\documentclass[aps,pre,12pt,notitlepage,superscriptaddress]{revtex4-2}

\usepackage{graphicx} 
\usepackage{amsmath}
\usepackage{amssymb}
\usepackage{xcolor}
\usepackage[normalem]{ulem}

\newcommand{\vecx}{{\bf x}}

\begin{document}

\title{Perspective: Measuring physical entropy out of equilibrium}

\author{Haim Diamant}
\affiliation{School of Chemistry and Center for Physics and Chemistry of Living Systems, Tel Aviv University, 69978 Tel Aviv, Israel}
\email{hdiamant@tau.ac.il}

\author{Gil Ariel}
\affiliation{Department of Mathematics, Bar-Ilan University, 52000 Ramat Gan, Israel}
\email{arielg@math.biu.ac.il}


\begin{abstract}
\setlength{\baselineskip}{16pt}
    Entropy is one of the key thermodynamic variables reflecting changes in the state of matter. Unlike other thermodynamic variables, it is well-defined also for nonequilibrium steady states through its relation to information. Applying this relation to physical systems is an ongoing challenge, as it requires knowledge of microscopic high-dimensional continuous distributions which is generally unattainable. A set of new approaches for the measurement of entropy in nonequilibrium steady or absorbing states have been developed and successfully applied to identify dynamic structures and transitions in diverse systems, ranging from jammed packings to swarming bacteria. We briefly review these approaches, emphasizing why applications to physical systems, including those out of equilibrium, is substantially different from the general statistical challenge of entropy estimation and inference. We point at promising current and future directions.
\end{abstract}

\maketitle

\begin{figure}
    \centering
    \includegraphics[width=0.44\linewidth]{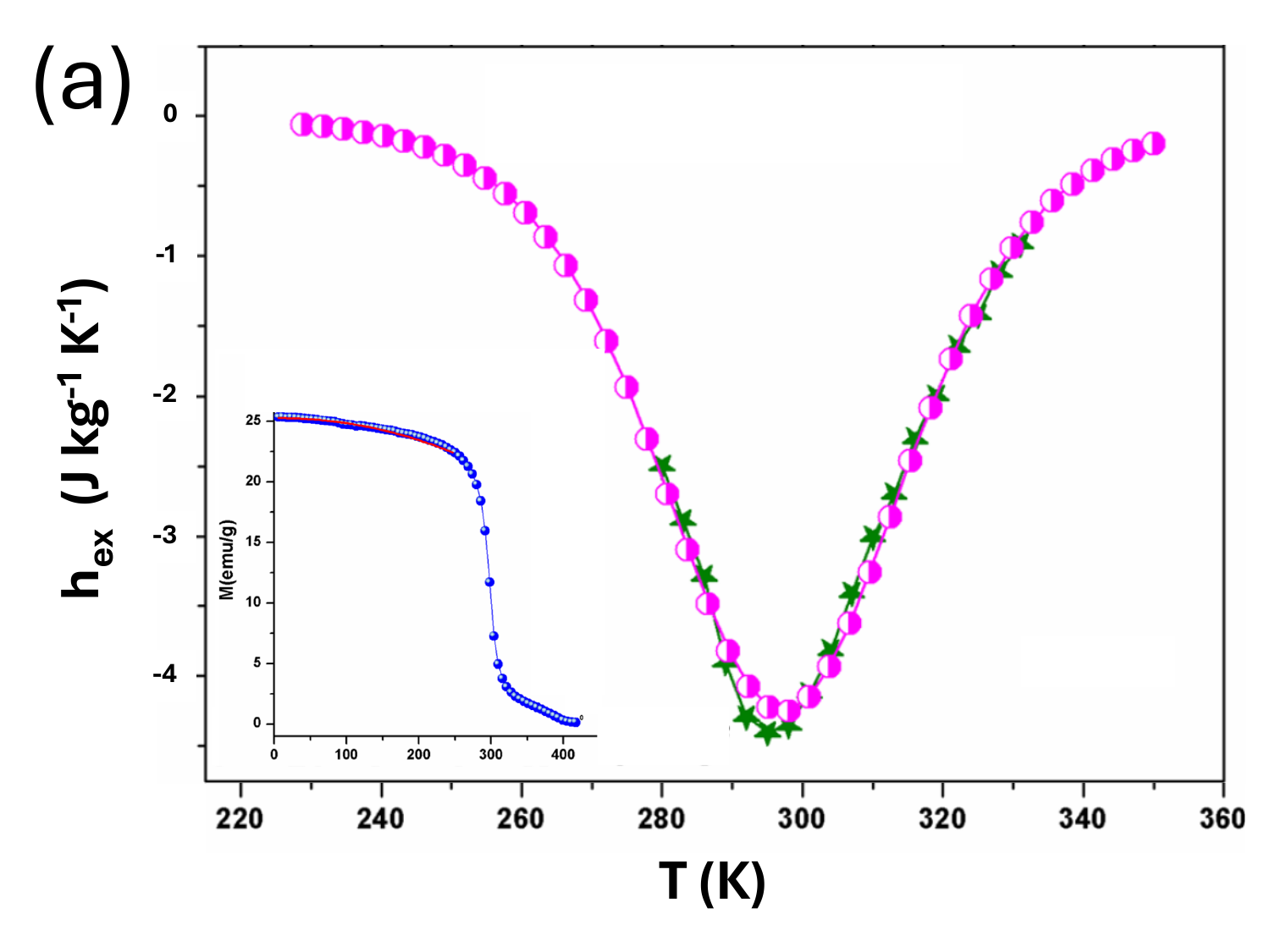}
    \includegraphics[width=0.54\linewidth]{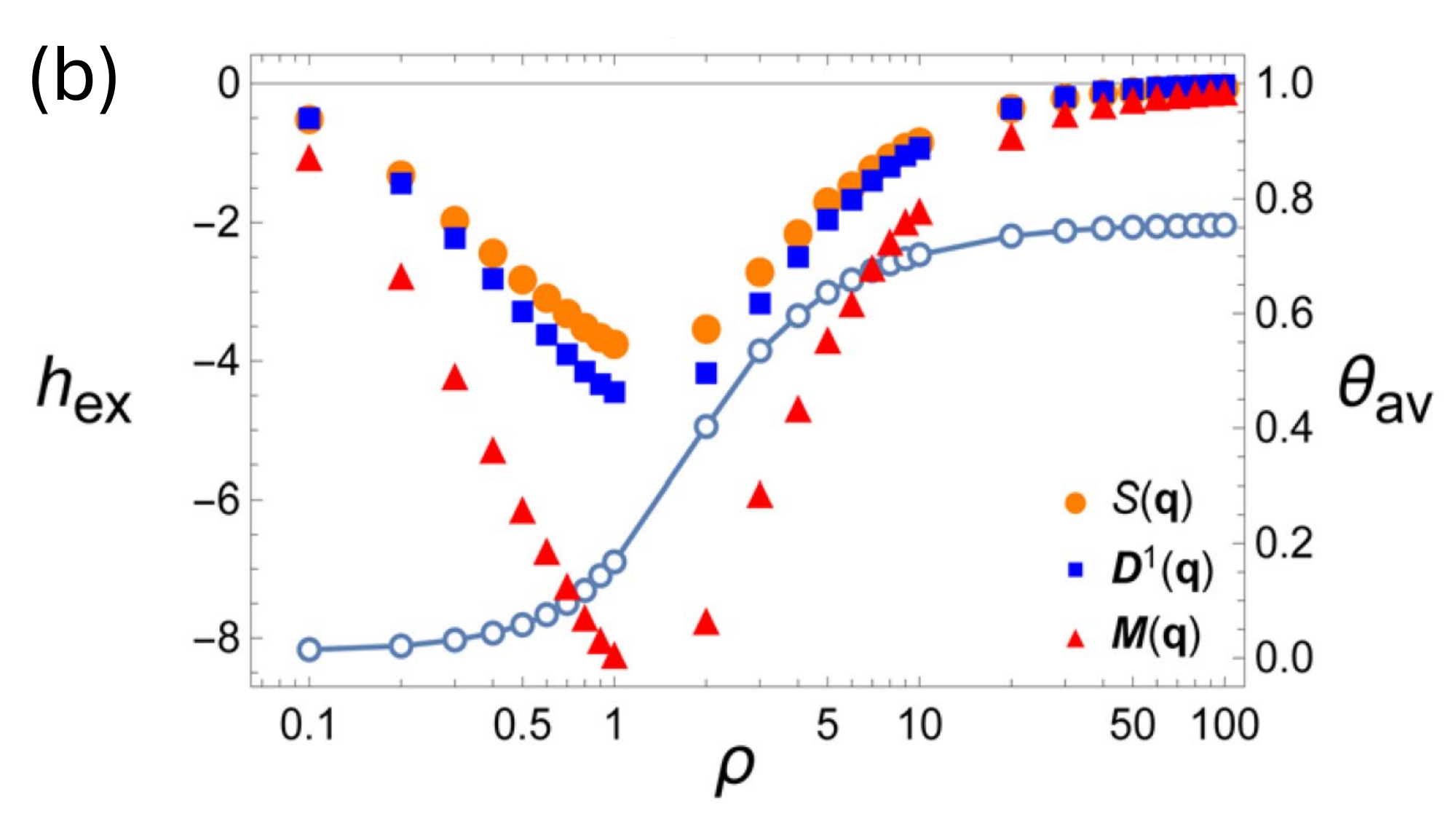}
    \caption{\setlength{\baselineskip}{14pt} Measuring entropy in and out of equilibrium. (a) Entropy change with temperature measured experimentally in a perovskite compound exhibiting a ferromagnet-to-paramagnet phase transition. The inset shows the temperature-dependent magnetization. Adapted from Ref.~\cite{Belkahla2019}. (b) Entropy change with density measured from simulations of the Vicsek model exhibiting a disorder-to-flock transition. The gray curve (with its values indicated on the right-hand vertical axis) shows the density-dependent orientational order parameter. Full colored symbols show different upper bounds for the entropy obtained using different correlation functions (see text). Adapted from Ref.~\cite{Sorkin2023phase}.}
    \label{fig:alloy_vicsek}
\end{figure}

Entropy is a thermodynamic state variable which distills the information contained in the statistics of a system's microstates into a single scalar. It reflects, in particular, changes in a material due to a phase transition. Figure~\ref{fig:alloy_vicsek}(a) shows measurements of the entropy change due to magnetization of a perovskite compound undergoing a ferromagnet-to-paramagnet phase transition as a function of temperature~\cite{Belkahla2019}. The transition is clearly reflected in a large drop of entropy. The entropy in this example was calculated from the derivative of the magnetic free energy with respect to temperature. Other means to measure such entropy changes use reversible heat flow, heat capacity, energy fluctuations, and the equation of state. All these methods apply only at thermal equilibrium. Out of equilibrium, the free energy, temperature, and pressure are generally not well-defined, the relation between entropy and energy fluctuations does not hold, and, importantly, entropy can by produced without heat flow \cite{Seifert2012,Seara2021,PelitiBook,Sorkin2024}. Yet, also a nonequilibrium system transitioning between two steady states should exhibit changes in entropy, reflecting the underlying changes in its distribution of microstates. Figure~\ref{fig:alloy_vicsek}(b) shows entropy changes measured in simulations of the Vicsek model~\cite{vicsek1995novel} undergoing a transition between disordered and flocking states as a function of density~\cite{Sorkin2023phase}. The entropy was measured using one of the methods described below. In this far-from-equilibrium active system, too, the transition entails a large entropy drop. The similarity between the two examples is visually evident, and so is the benefit from entropy measurement for pinpointing the two transitions. Resolving the entropy changes in the active example, however, requires methods for entropy measurement out of equilibrium, which is the subject of this Perspective.

The starting point of all these methods is the Shannon informational entropy,
\begin{equation}
    H[P_s] = -\sum_s P_s \ln P_s,\ \ \ \ 
    H[p(\vecx)] = -\int d\vecx\, p(\vecx) \ln[\nu p(\vecx)].
\label{Shannon}
\end{equation}
We have given the entropy for a probability distribution $P_s$ of discrete states $s$ and a probability density $p(\vecx)$ of a continuous phase space $\vecx$. The normalization constant $\nu$ takes care of units, and we take the Boltzmann constant to be $1$. Equation~(\ref{Shannon}) defines entropy in terms of the distribution of microstates alone. In particular, it does not make any physical assumptions such as conservation laws and thermodynamic equilibrium. To apply Eq.~(\ref{Shannon}) directly, one needs good sampling of the many-variables $P_s$ or the infinitely-many-variables $p(\vecx)$. The dimension of the full phase-space is typically extremely high, and samples taken consecutively in experiments or simulations may be correlated. This is very far from a finite string of bits, making various statistical computer-science techniques of entropy calculation highly inefficient. Thus measuring the entropy of physical systems is an enormous challenge. In particular, traditional Monte-Carlo methods fail, as demonstrated by the following simple example.

\begin{figure}
    \centering
    \vspace{-2cm}
    \includegraphics[width=0.9\linewidth,trim={0 2cm 8.5cm 0},clip]{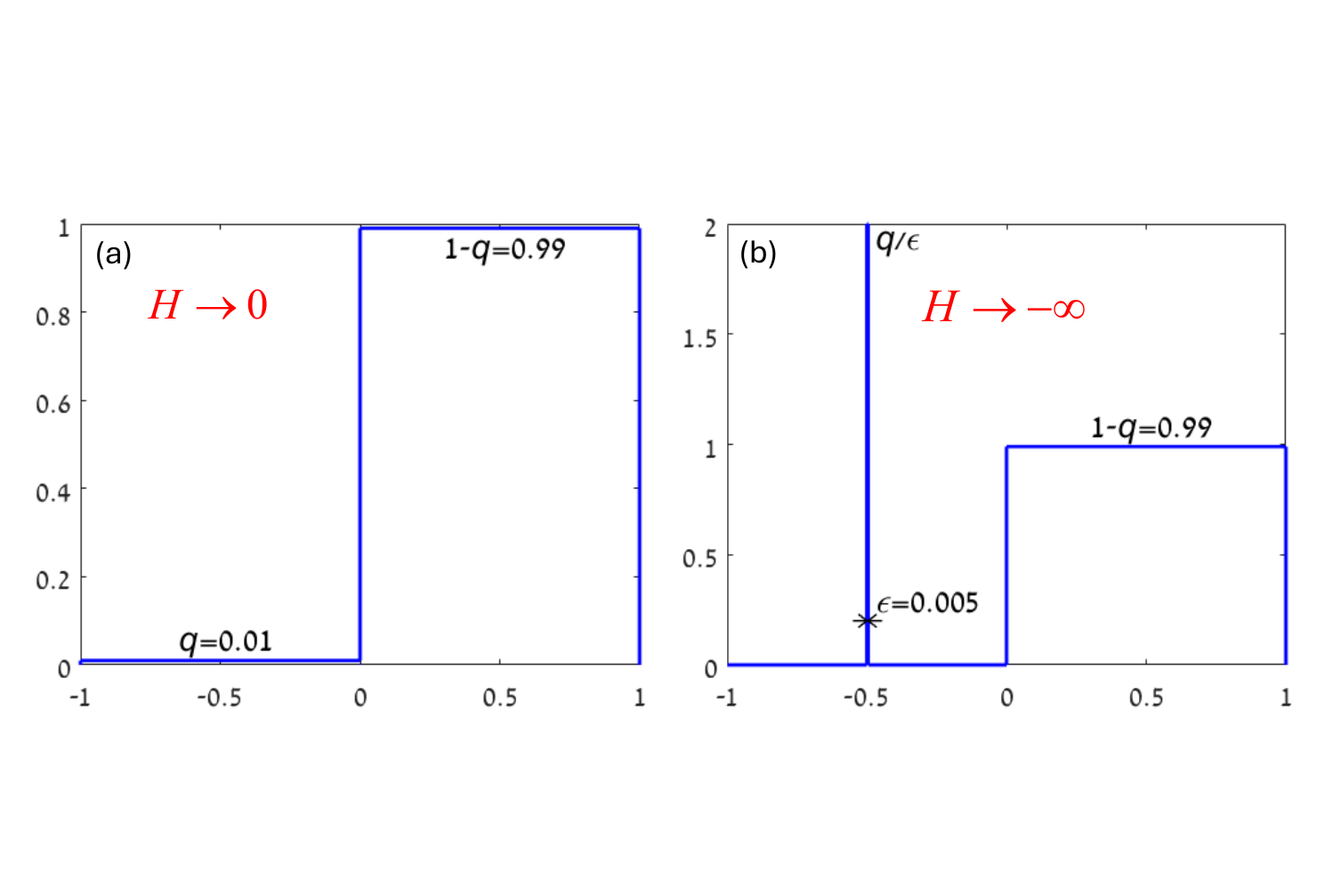}
    \vspace{-2.0cm}
    \caption{\setlength{\baselineskip}{14pt} Demonstration of the difficulty in estimating the entropy of systems with continuous probability distributions. Panels (a) and (b) show two probability densities of a scalar variable. For small $q$, the entropy of (a) tends to zero. In sharp contrast,  with $\epsilon \ll q$, the entropy of  (b) can become boundlessly negative. Distinguishing between the two cases by sampling requires $\sim 1/q$ samples. A system with an $M$-dimensional phase space would require a prohibitively large number of $q^{-M}$ samples.}
    \label{fig:sampling}
\end{figure}

Consider two scalar random variables, distributed over $[-1,1]$ with the two densities depicted in Fig.~\ref{fig:sampling}. In both cases the probability to sample a value in $[-1,0]$ is $q$, and in $[0,1]$ it is $1-q$. While the distribution in (a) is uniform in each half ($[-1,0]$ and $[0,1]$), in (b) it is uniform on the positive half and concentrated (uniformly) in a narrow segment of width $\epsilon$ on the negative half. For $q \ll 1$, sampling both distributions will give similar results\,---\,an overwhelmingly large fraction of the samples will be positive. Yet, the entropy of the two cases could not be more different. From Eq.~(\ref{Shannon}), 
the entropy of (a) is $H_{\rm a}=-q(\ln q - 1) + O(q^2) \xrightarrow[q\to 0]{} 0$, whereas the entropy of (b) is $H_{\rm b} = -q \ln(q/\epsilon) + O(q) \xrightarrow[\epsilon\to 0]{} -\infty$. Thus, while the entropy properly captures the sharp difference between the two cases, insufficient random sampling will miss this distinction and also wrongly estimate $H_{\rm b}$. 
For a physical system whose phase space is $M$-dimensional, a similar problem will occur in each dimension, requiring  $\sim q^{-M}$ samples just to decide whether we are in one case or the other. This difficulty does not depend on the estimator used to calculate the entropy. Hence, when dealing with physical systems of even moderate phase space, we are always in a severely under-sampled regime.

Any estimation method, when applied to a high-dimensional physical system, will implicitly involve some assumptions on the underlying distribution. This is where physics comes to help. If ample sampling of phase space is impossible, one may devise inference methods that make additional, physically motivated assumptions. We will give examples below. 
In addition, in certain cases the method may be successful in detecting transitions, not because it accurately estimates the entropy, but because the underlying assumptions {\em succeed or fail differently} in different regimes. For example, a method (such as the CID method mentioned below), may be sensitive to short-range correlations compared to long-range ones. As a result, a transition involving a sharp change in the range of correlations will be manifested in changes of the estimated entropy, even if the actual entropy values may be inaccurate.   

We now briefly survey several approaches proposed recently for entropy measurement in physical systems. This Perspective is far from a comprehensive review. We have picked only a few works which we find of particular potential.

\vspace{12pt}
{\it Compression-based methods.}

Entropy estimation via compression rests on a core insight from Shannon's source coding theorem, stating that the entropy of a source sets a lower bound on the average number of bits needed to encode its output. A good compressor approaches that bound, so the compressed length of a sequence serves as an approximation of its entropy. Computational methods, such as context-tree weighting, have exploited these properties since the 1980's \cite{willems2002context}. The works of Refs.~\cite{Martiniani2019,Avinery2019} suggested, independently, to use off-the-shelf lossless compression algorithms for obtaining the resulting approximation for the entropy, termed computable information
density (CID), to study physical ensembles. 

Such compression methods are simple to implement, highly generic, and can be applied ``blindly", without prior knowledge of the physical system. They perform surprisingly well in diverse situations. In particular, they were successfully applied to various far-from-equilibrium transitions,  including transitions from active to static absorbing states in schematic dynamic models~\cite{Martiniani2019}, motility-induced phase separation of active Brownian particles~\cite{Martiniani2019}, protein folding~\cite{Avinery2019}, and the flocking transition in the Vicsek model ~\cite{Cavagna2021}. The method was used also to extract correlation lengths from the CID of progressively ``diluted"  spatial configurations~\cite{Martiniani2020}.

However, compression-based methods depend on additional extrinsic factors, such as the transformation of the data into a linear string and the scaling of compression output to entropy~\cite{Avinery2019}. The former issue may hamper the identification of long-range correlations~\cite{Zu2020}. As they still depend on a sufficient number of samples, they also show sensitivity to system size~\cite{ArielDiamant2020}.

\vspace{12pt}
{\it Entropy bounds from correlation functions.}

Because of the difficulties explained above, it makes sense to relate information-theoretic quantities to integrated observables which are more accessible experimentally and numerically. One important family of such observables are correlation functions. This approach can be traced back eight decades to the work by Green, who used equilibrium liquid theory to relate the entropy of a fluid to the particles' pair distribution function $g({\bf r})$
\cite{Green1947,Laird1992},
\begin{equation}
  h - h_{\rm id} = -\frac{\rho}{2}
  \int d{\bf r} \left[ g({\bf r}) \ln g({\bf r}) + 1 - g({\bf r}) \right],
\label{eq:hG}
\end{equation}
where $h=H/N$ is the entropy per particle, $h_{\rm id}$ its value for uncorrelated particles, and $\rho$ the mean particle density. 

A maximum-entropy approach has recently been used to derive upper entropy bounds with spatial two-point correlation functions as constraints \cite{ArielDiamant2020,Sorkin2023entropy,Sorkin2023phase}. Applicable both in and out of equilibrium, the method yields the maximum entropy that a system can have given certain correlation functions. This approach bypasses the problems of sampling, sample discretization, system size, and the possible presence of long-range correlations. On the other hand, the bound may not be tight. The bound is given in the form of a functional of the Fourier-transformed spatial correlation functions arising from any microscopic degrees of freedom. The simplest example is the upper bound obtained from the position correlation function, i.e., the structure factor, which is an alternative to Green's Eq.~(\ref{eq:hG})~\cite{ArielDiamant2020}. More generally,
\begin{equation}
    h - h_{\rm id} \,\leq\, \frac{1}{2\rho} \int \frac{d{\bf q}}{(2\pi)^d}\, {\rm tr} \left[ \ln {\cal C}({\bf q}) - I + {\cal C}({\bf q}) \right],
\label{Hcorrelation}
\end{equation}
where ${\cal C}({\bf q})$ is a matrix including all known two-point correlations and cross-correlations, as a function of wavevector ${\bf q}$, $I$ is the corresponding unit matrix, and $d$ is the system's dimensionality.

Importantly, when several correlation functions are known, comparing the different bounds obtained from them allows for the identification of the degrees of freedom governing a certain physical regime. Figure~\ref{fig:alloy_vicsek}(b) gives an example using simulations of the Vicsek model. Orange dots, blue squares, and red triangles show the entropy bounds obtained, respectively, from the correlation functions of particle positions, orientations, and all measured correlations (positions, orientations, and their cross-correlation). We see that the largest contribution to the entropy change accompanying the flocking transition comes from the orientational degree of freedom. In addition, the entropy change arising from all the correlations is to a good approximation equal to the sum of the contributions from positions and orientations, implying that their cross-correlation is not a crucial ingredient of the transition. These conclusions change for different magnitudes of noise used in the simulations, revealing the different mechanisms governing the flocking transition under different conditions~\cite{Sorkin2023phase}. The same approach was applied to transitions in swarming bacteria~\cite{Sorkin2023phase} and jammed packings of a binary mixture of droplets~\cite{Sorkin2023entropy}. For the Vicsek model, the entropy was found to be significantly more sensitive than the order parameter for detecting the known phase transition, especially for a small number of particles. For bacteria, it revealed a new transition which had not been recognized before, demonstrating a direct benefit from entropy measurement.

\vspace{12pt}
{\it Parametrization using machine learning.}

Neural networks provide powerful tools for approximating functions and densities. These have been used to estimate mutual information and entropy generally in data-science and machine-learning applications~\cite{belghazi2018mine}. 
Most machine-learning approaches rely on some form of the Donsker-Varadhan duality formula \cite{donsker1975asymptotic}, which expresses information-theoretic quantities as a variational problem. For example, the Kullback-Leibler divergence between two probability measures $P({\bf X})$ and $Q({\bf X})$ over a space ${\bf X}$ is given by
\begin{equation}
D_{KL}(P \| Q) = \sup_{T: {\bf X} \to \mathbb{R}} [\langle T \rangle_P - \log \langle e^T \rangle_Q],
\label{eq:DV}
\end{equation}
where $\langle \cdot \rangle_P$ denotes averaging with respect to $P$, and similarly for $Q$. The maximum is taken over an appropriate space of functionals $T({\bf X})$, e.g., bounded functionals. 

For physical systems, the entropy was calculated in Ref.~\cite{Nir2020} through an iterative division of the system into progressively smaller parts and estimating the mutual information between them using machine learning to parametrize the space of functionals. The method was used to study the jamming transition in a binary mixture of soft disks.
In Ref.~\cite{gelman2024nonequilibrium} an off-the-shelf machine-learning estimator was applied to estimate the probability density of the system's configurations, and hence the entropy. The method was used to study a dynamic transition in a 2D Ising model driven by an oscillating magnetic field.

\vspace{18pt}
In what follows we focus on new directions in entropy measurement which we find of particular promise.

\vspace{12pt}
{\it Direction 1: Bounds from kinetics.}

Steady-state entropy, which is a static property of the system's time-independent microstate distribution $p({\bf x})$, is not expected to be related to the system's kinetics, i.e., the way it transitions from one microstate to another over time. Many choices of kinetics may lead to the same entropy. This freedom does not rule out, however, the existence of constraints limiting the physically possible kinetics. Near equilibrium, Kubo relations and Onsager relations are classical examples.

The entropy has recently been proved to obey the following inequality~\cite{Sorkin2023kinetics}:
\begin{equation}
    H \leq -\int d{\bf x}' \, p({\bf x}') \int d{\bf x} \, w({\bf x}|{\bf x}';\tau)\ln[\nu w({\bf x}|{\bf x}';\tau)],
\label{propagator}
\end{equation}
which holds for Markovian dynamics at steady state arbitrarily far from equilibrium.
Here, $w({\bf x}|{\bf x}';\tau)$ is the transition probability (propagator) from microstate ${\bf x}'$ to microstate ${\bf x}$ during the system's relaxation time $\tau$, and $\nu$ is a normalization constant. The right-hand side of Eq.~(\ref{propagator}) can be interpreted as the Shannon entropy of the propagator, averaged over initial conditions. 

Equation (\ref{propagator}) is a good starting point for deriving more useful entropy bounds at the expense of reducing the bound's tightness. For example, assuming that $w$ is a product of single-particle Brownian propagators produces an entropy bound from the particles' self-diffusion coefficient $D$ and the relaxation time $\tau$ \cite{Sorkin2023kinetics},
\begin{equation}
    h - h_0 \,\leq\, \frac{d}{2} \ln \frac{D\tau}{D_0\tau_0},
\label{kinetics}
\end{equation}
where the subscript $0$ indicates values in a certain reference state. Thus, an upper bound for the system's entropy can be obtained from the kinetic properties $D$ and $\tau$, which are quite readily measured. Equation~(\ref{kinetics}) is reminiscent of empirical entropy scaling relations, which have been used for many years~\cite{Rosenfeld1977,Dyre2018}.

It should be possible to use Eq.~(\ref{propagator}) to derive additional upper bounds based on other kinetic coefficients such as electrical conductivity, heat conductivity, viscosity, etc. As in the case of the bounds from correlation functions, comparing the different bounds may allow us to identify the dominant physical mechanism underlying changes of entropy in nonequilibrium transitions.

\vspace{12pt}
{\it Direction 2: Quantum entropy.}

The entropy of a quantum system can be found from its density matrix $\hat\rho$ via the von Neumann entropy~\cite{vonNeumannBook},
\begin{equation}
    H = -{\rm Tr}\, (\hat\rho \ln \hat\rho).
\label{vonNeumann}
\end{equation}
Being a generalization of the Shannon entropy [Eq.~(\ref{Shannon})] to quantum systems, the von Neumann entropy is a measure of the degree of mixing of the system's quantum state; it is zero for a pure state and maximum for a mixture of equally weighted states in the system's Hilbert space. At thermal equilibrium the von Neumann entropy yields the thermodynamic entropy. For example, maximizing it under the constraint of a given expectation value of the Hamiltonian gives the canonical Boltzmann distribution, etc. Like Shannon's formula, Eq.~(\ref{vonNeumann}) can be applied just as well to the density matrix of a nonequilibrium steady state. 

Measuring the von Neumann entropy should be as helpful for quantum systems as measuring the Shannon entropy is for classical ones. The development of methods for this purpose, similar to the ones described above, seems feasible. For example, one should be able to derive bounds on the von Neumann entropy from spatial correlation functions. Moreover, for a given physical system, the von Neumann entropy minimizes the Shannon entropy obtained from the probability distribution of possible measurement results~\cite{WildeArxiv}. Thus, apart from the extension to quantum systems, the von Neumann entropy may be useful for improving the entropy upper bounds obtained for the same physical system in the classical limit. Another direction is to generalize the entropy estimation from the mutual information of iterative slicing of the system, mentioned above, to quantum mutual information~\cite{WildeArxiv}.

\vspace{12pt}
{\it Direction 3: Entropy production.}

While steady-state entropy arises from the statistics of the system's microstates, irrespective of their time sequence, entropy production (EP) has to do with the stochastic trajectories connecting these microstates over time. EP is directly related to energy dissipation, indicating the degree of irreversibility of the process that the system undergoes~\cite{Seifert2012,Sorkin2024,PelitiBook}. For overdamped dynamics it can be obtained from the statistics of trajectories as~\cite{PelitiBook},
\begin{equation}
    {\rm EP} = D_{KL} \{{\rm Pr}[\overrightarrow{\bf x}(t)]\, \|\, {\rm Pr}[\overleftarrow{\bf x}(t)] \},
\label{EP}
\end{equation}
where $\overrightarrow{\bf x}(t)$ denotes a time sequence of microstates, and $\overleftarrow{\bf x}(t)$ the inverted sequence of the same microstates. As the distribution of trajectories is even harder to sample in experiment or simulation than that of the microstates, the essential difficulty in direct estimation of entropy, described in the beginning, worsens tremendously for EP.

A variety of methods have been developed for inferring EP from empirical measurements of nonequilibrium stochastic systems.
Most works only attempt to derive lower bounds for EP obtained by coarse-graining. Such methods typically rely on thermodynamic inequalities such as
the thermodynamic uncertainty relation \citep{Barato2015,Gingrich2016,Li2019,Manikandan2020}, which bound EP in terms of the mean and variance of lower-dimensional probability currents.
Other techniques relate EP to the statistics of waiting times \citep{Skinner2021a,VanderMeer2022,Harunari2022,Martinez2019}, observed transitions \citep{Bisker2017,Polettini2017,VanderMeer2023}, and counting observables \citep{Pietzonka2024}. Many of them are designed to infer the underlying dissipation from a small number of coarse-grained observables \citep{Ehrich2021,Nitzan2023,Knotz2024}.
The work of Ref.~\citep{Seara2021} presented a bound based on the cross-spectral density matrix within a Gaussian approximation. In Ref.~\citep{Aguilera2026} the EP was formulated as a (convex) maximization problem [along the lines of Eq.~(\ref{eq:DV})], with measured observables such as spatiotemporal correlation functions as constraints. 

A few model-free methods for measuring local EP have been suggested. A compression-based method based on cross-parsing complexity was introcued in Ref.~\citep{Ro2022}. The work in Ref.~\citep{Kim2020} used an estimator, in which a neural network learns a function whose output approximates the log-ratio of forward to reverse transition probabilities [see Eq.~(\ref{EP})]. The procedure was proved to recover the stochastic EP at optimality.
A deep-learning framework was developed in Ref.~\citep{Boffi2024} to estimate local probability currents and EP. 
Clearly, simple, possibly analytic methods for EP measurement, which will be useful for experiments and simulations, remain a central challenge.

\vspace{12pt}
In conclusion, the measurement of entropy in nonequilibrium physical systems reveals new mechanisms and provides new insights, in particular, in systems undergoing nonequilibrium transitions. The data used as input to the various entropy-estimation methods\,---\,sampled configurations and trajectories, image files, measured correlations\,---\,already contain the available information; yet, the distillation of the data into the single scalar $H$ is found to be of great benefit. The directions sketched above have the potential for important advances in the measurement of entropy and entropy production in complex physical systems. Such advances will hopefully lead to new discoveries concerning dynamic and arrested states of disordered, active, and living matter.

\vspace{12pt}
\begin{acknowledgments}
We acknowledge support from the Israel Science Foundation under Grant No.\ 1611/24.
\end{acknowledgments}

\bibliographystyle{apsrev4-2}
\bibliography{references}

\end{document}